\documentclass[prb,twocolumn,floatfix,showpacs,preprintnumbers,superscriptaddress,amsmath,amssymb]{revtex4}
\usepackage[dvips]{graphics}

\def\pfhv{\hat{\mbox{\boldmath$p$}}_{\!\!{\scriptscriptstyle F}}}
\def\vp{{\bf p}}
\def\vv{{\bf v}}
\def\vy{{\bf y}}
\def\vz{{\bf z}}
\def\vH{{\bf H}}

\usepackage{graphicx}

\begin{document}

\title{Subharmonic gap structure in d-wave superconductors}

\author {A. Poenicke} 
\affiliation{Institut f\"ur theoretische Festk\"orperphysik, Universit\"at Karlsruhe, 
76128 Karlsruhe, Germany}

\author {J.C. Cuevas}
\affiliation{Institut f\"ur theoretische Festk\"orperphysik, Universit\"at Karlsruhe, 
76128 Karlsruhe, Germany}

\author {M. Fogelstr\"om} 
\affiliation{Institute of Theoretical Physics, Chalmers University of
Technology and G\"oteborgs University, S-41296 G\"oteborg, Sweden}

\date{\today}

\begin{abstract}
We present a self-consistent theory of current-voltage characteristics of d-wave/d-wave 
contacts at arbitrary transparency. In particular, we address the open problem of the 
observation of subharmonic gap structure (SGS) in cuprate junctions. Our analysis shows 
that: (i) the SGS is possible in d-wave superconductors, (ii) the existence of bound 
states within the gap results in an even-odd effect in the SGS, (iii) elastic scattering 
mechanisms, like impurities or surface roughness, may suppress the SGS, and (iv) in the 
presence of a magnetic field the Doppler shift of the Andreev bound states leads to the 
splitting of the SGS, which is an unambiguous fingerprint of d-wave superconductivity.
\end{abstract}


\maketitle

Although the d-wave scenario is emerging as the new paradigm in superconducting cuprates
\cite{Scalapino1995,Harlingen1995,Tsuei2000}, there is still a lacking consensus with 
regards to current transport in SIS junctions \cite{Kashiwaya2000}. By now it is clear that 
tunneling in unconventional superconductors is a phase-sensitive technique. An important 
consequence of an order parameter showing a sign change in different momentum directions 
is the formation of Andreev bound states at zero energy confined to the
surface \cite{Buchholtz1981,Hu1994}. One of the consequences of the presence of these 
bound states is the appearance of a zero-bias conductance peak in the current-voltage 
characteristics. This peak has been observed in different experiments using SIN-type 
junctions \cite{Kashiwaya1995} and SIS grain boundaries \cite{Alff1998}. However, there 
is still a controversy related to the observation of the subharmonic gap structure (SGS)
in cuprate SIS junctions. Devereaux and Fulde suggested that the Andreev scattering in SNS 
contacts could be used to identify the symmetry of the order parameter \cite{Devereaux1993}.
Based on their analysis the SGS revealed in different experiments in YBCO contacts were interpreted 
as an evidence of the existence of a well-defined nonvanishing gap \cite{Polturak1993,Ponomarev1995}.
After more detailed but non-self consistent analysis, different authors have concluded that the SGS 
are weak in d-wave superconductors, since averaging over the anisotropic gap washes out any 
prominent features \cite{Hurd1997,Lofwander2001}. However, there are also more recent experimental 
reports of SGS in YBCO edge Josephson junctions \cite{Engelhardt1999,Nesher1999}, and even 
in the c-axis tunneling of different cuprate contacts \cite{Ponomarev2001}. Auerbach and Altman 
proposed an alternative interpretation of the appearance of pronounced SGS in 
Ref.~[\onlinecite{Nesher1999}], claiming that this structure is an indication of magnon tunneling 
that can be explained in the context of the SO(5) theory \cite{Auerbach2000}. 

In this Communication we present a self-consistent theoretical analysis of the ac Josephson effect 
in d-wave superconductors. Using quasiclassical methods we determine the local electronic 
properties of the superconductors in the interface region. This includes effects on the order 
parameter (OP) profile and on the local density of states (DOS) by pair breaking caused 
both by quasiparticle scattering off the interface and off homogeneously distributed impurities 
in the crystals \cite{Matsumoto1995,Buchholtz1995,Fogelstrom1997,Barash1997,Poenicke1999,
Fogelstrom-preprints}. The I-V characteristics of the d-wave/d-wave SIS-junction are 
computed using the local surface Green's functions containing all relevant information of
superconducting electrodes and solving the appropriate boundary conditions for a point contact 
\cite{Cuevas2001}. We shall show that, not only does the SGS survive after averaging for all 
crystal misorientations, but there are qualitatively new features in the I-V characteristics 
that can only be captured by a truly self-consistent calculation. For instance, there is an 
even-odd effect in the SGS which originates from processes connecting the zero-energy states
with Andreev states formed close to the gap edges. Such an even-odd effect in the differential 
resistance has been reported in YBCO edge-junctions by Nesher and Koren \cite{Nesher1999}. These 
states are missed in a non-self consistent calculation where the suppression of the OP close to 
the interface is neglected. We also show that the SGS in d-wave contacts can be tuned with a 
magnetic field. The Doppler shift of the Andreev bound states \cite{Fogelstrom1997,Aprili1999} leads 
to a very peculiar splitting of the SGS, which provides a clear signature of d-wave superconductivity.

In the limit of not too low interface transparency, $D\ge 0.1$, the main feature of the I-V 
characteristics of conventional SIS contacts is the appearance of SGS, which consists of a series 
of maxima in the conductance at voltages $eV = 2\Delta/n$, where $\Delta$ is the superconducting
gap and $n$ is an integer number. This peculiar structure is due to the occurrence of multiple 
Andreev reflections (MAR) and by now well established both theoretically \cite{Averin1995} and 
experimentally \cite{Scheer1997} in the case of s-wave superconductors. In this work, our goal is 
to extend the analysis of the SGS to the case of superconducting cuprates. For this purpose, we 
consider a voltage biased contact, consisting of two $d_{x^2 - y^2}$ superconductors separated by 
a single interface of arbitrary transparency. The order parameter on side $i$, $i=L,R$, is rotated 
by $\alpha_i$ with respect to the surface normal, and we denote junction type by the relative
crystal orientations as $d_{\alpha_L}$-$d_{\alpha_R}$. There are several experimental realizations 
of this system, among which the bicrystal grain-boundary junctions are ideal examples 
\cite{Alff1998}. We carry out the calculation of the current following the formulation 
introduced by two of the authors in Ref.~[\onlinecite{Cuevas2001}], and refer the reader to this 
work for all the technical details. In the case of a constant bias voltage, $V$, one can show that
the current oscillates in time with all the harmonics of the Josephson frequency, i.e. $I(t) = 
\sum_m I_m e^{im \phi(t)}$, where $\phi(t)=\phi_0+(2eV/\hbar)t$ is the time-dependent superconducting 
phase difference. We concentrate ourselves in the analysis of the dc current, denoted from now on as $I$. 
Furthermore, we assume that the interface conserves the momentum of the quasiclassical trajectories, 
which allows us to write the current as a sum over independent trajectory contributions:
$I = \frac{1}{2} \int^{\pi/2}_{-\pi/2} d\pfhv \; I(\pfhv) \cos(\pfhv)$, where $\pfhv$ defines
the Fermi surface position. For the angular dependence of the transmission coefficient we use the 
expression $D(\pfhv) = D \cos^2(\pfhv)/ [1-D\sin^2(\pfhv)]$, resulting from a $\delta$-like potential. 
Here $D$ is the transmission for the trajectory perpendicular to the interface.

\begin{figure}[h]
\begin{center}
\includegraphics[width=\columnwidth,clip=]{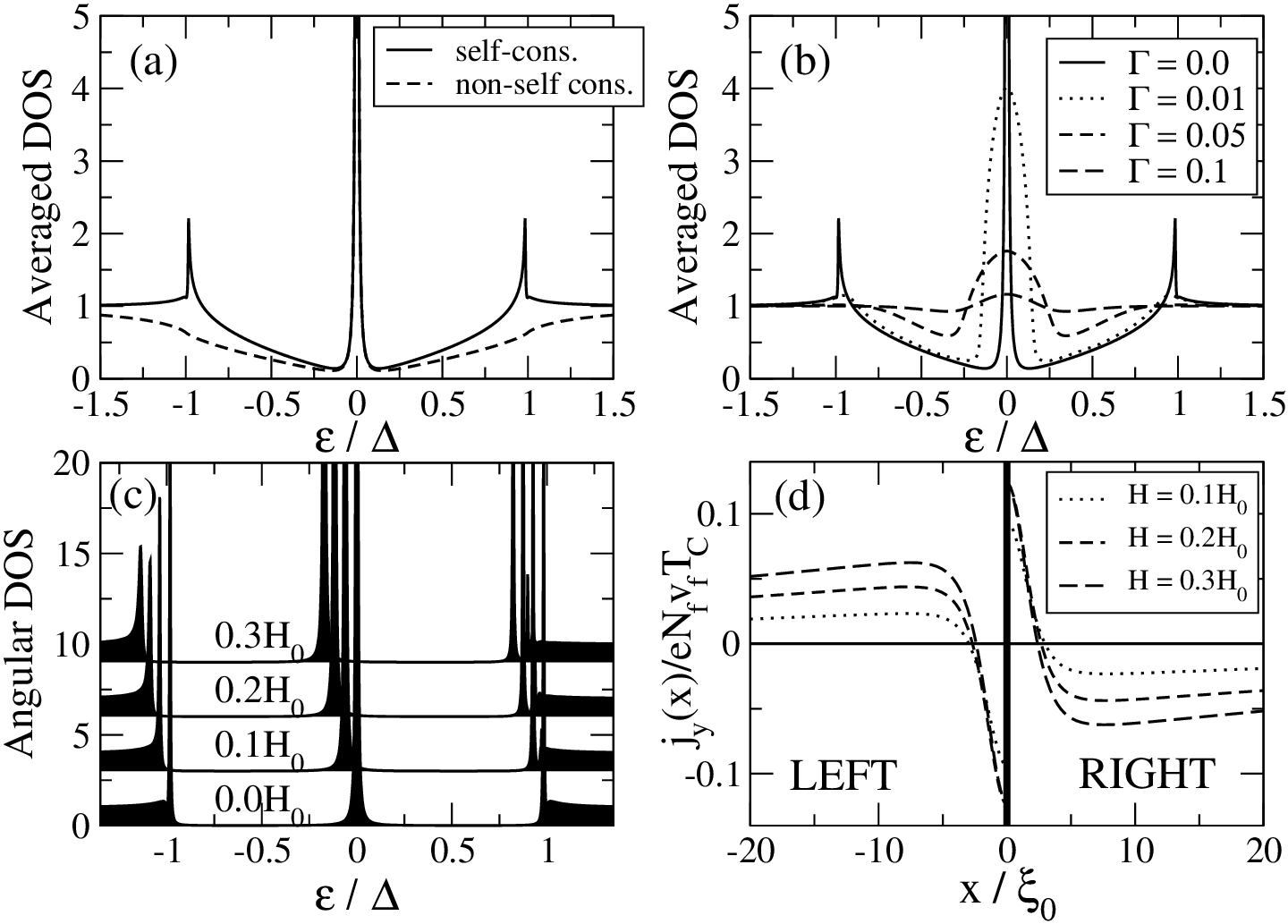}
\vspace*{-0.3truecm}
\caption{(a) Angle-averaged local DOS at the interface for a $45^o$ misorientation (clean case):
self-consistent and non-self consistent. (b) Local DOS for a $45^o$ misorientation for different 
values of the bulk-impurity scattering rate $\Gamma$ (Born scatterers), measured in units of 
$2\pi T_C$, where $T_C$ is the critical temperature in the clean case. $\Delta$ is the maximum 
bulk gap for the clean superconductor. (c) Angular resolved DOS for a trajectory $\pfhv=45^o$ for 
different magnetic fields.  All the DOS are normalized to the normal state DOS, $N_f$.
(d) The current density parallel to the interface, $j_y$, at different applied fields for a 
$d_{\pi/4}$-$d_{-\pi/4}$ contact. The distance to the interface, $x$, is normalized with the
clean coherence length, $\xi_0$.}
\label{Spectra}
\end{center}
\end{figure}

\begin{figure}[h]
\begin{center}
\includegraphics[width=0.71\columnwidth,clip=]{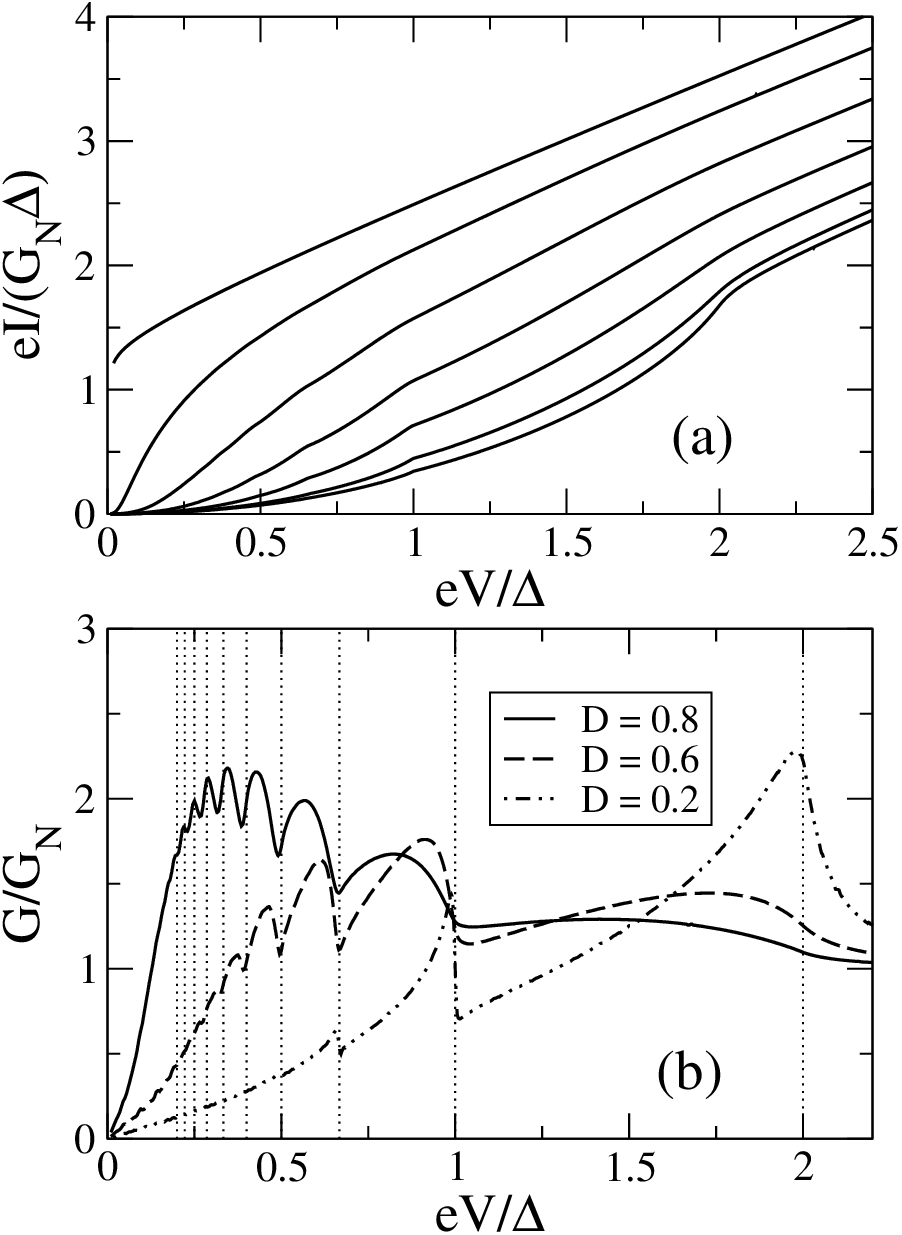}
\vspace*{-0.3truecm}
\caption{$d_0$-$d_0$ contact in the clean case: (a) I-V characteristics for different 
transmissions, from bottom to top $D=0.1,0.2,0.4,0.6,0.8,0.95,1.0$. (b) Differential 
conductance normalized by the normal state conductance $G_N$. The dotted vertical lines 
indicate $eV = 2\Delta/n$.}
\label{0-0}
\end{center}
\end{figure}

Let us start by analyzing the case of a symmetric $d_0$-$d_0$ junction in the clean
limit. In this case, the order parameter is constant up to the surface, and there
are no bound states for any trajectory. The I-V characteristics of a single trajectory,
$I(\pfhv)$, coincide with those of isotropic s-wave superconductors, and exhibit a
pronounced SGS at $eV=2\Delta(\pfhv)/n$ [\onlinecite{Averin1995}]. Since the different trajectories 
see different gaps, the relevant question is whether the SGS survives after averaging. The 
answer can be seen in Fig.~\ref{0-0}, where both the current and differential conductance
are shown for different transmissions. One can clearly see a SGS at $eV = 2\Delta/n$,
although it is more rounded than in the s-wave case. Notice also that the SGS are
better defined as minima in the conductance at the subharmonic voltages, i.e. as
maxima in the differential resistance. At this point a natural question arises:
why is the maximum gap the energy scale revealed in the SGS? The idea is that the
jump of the current at the opening of a new MAR scales with the gap of the
corresponding trajectory. Therefore, the trajectories with the largest gaps
dominate the contribution to the SGS.

\begin{figure}[h]
\begin{center}
\includegraphics[width=.71\columnwidth,clip=]{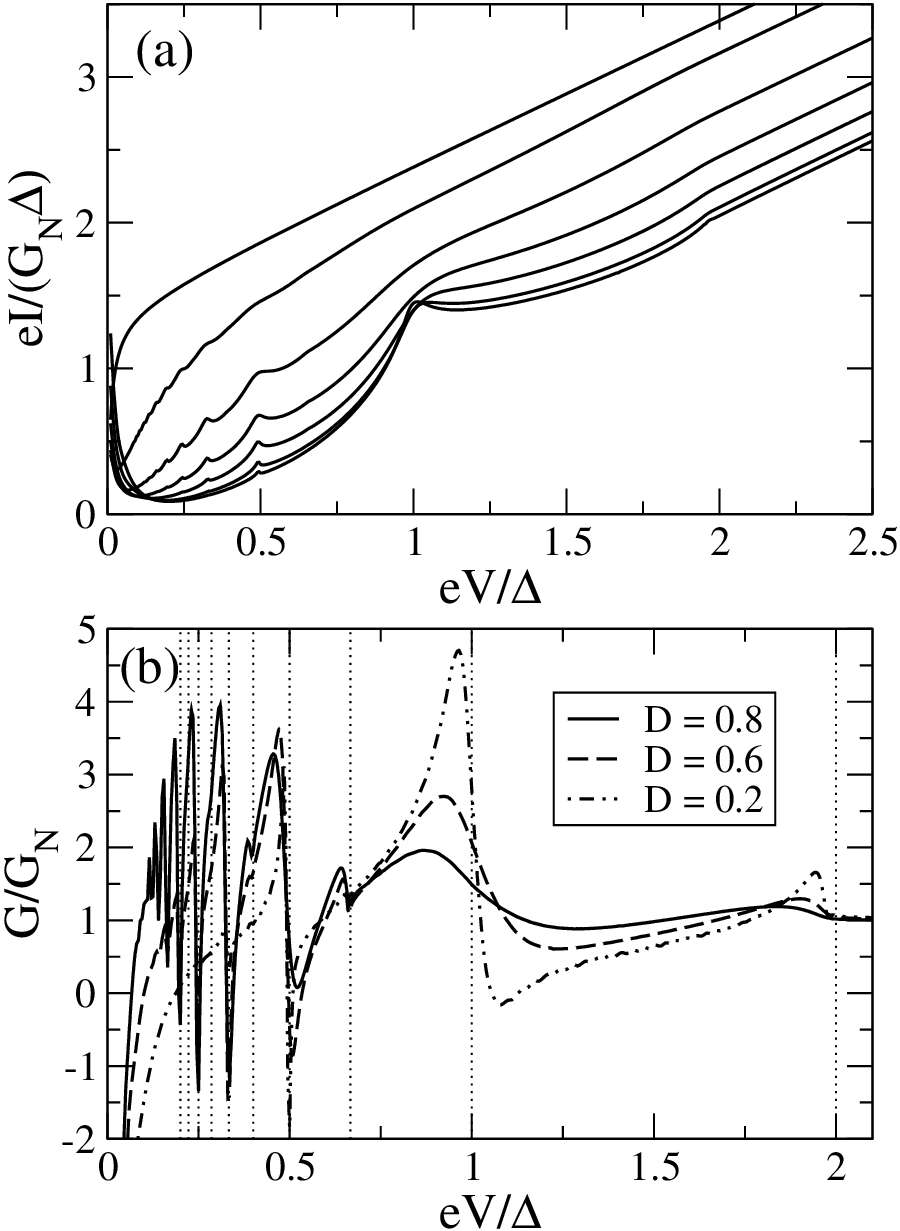}
\vspace*{-0.3truecm}
\caption{$d_{\pi/4}$-$d_{-\pi/4}$ contact in the clean case: (a) I-V characteristics 
for the same transmissions as in Fig.~2a. (b) Differential conductance.}
\label{45-45}
\end{center}
\end{figure}

Let us now consider the case of a $d_{\pi/4}$-$d_{-\pi/4}$ junction. In this case,
assuming specular quasiparticle scattering at the interface, an Andreev bound state
forms at zero energy for every trajectory \cite{Hu1994}. This implies
that the surface acts as a pair-breaker \cite{Buchholtz1995} and the gap is
depressed in the vicinity of the interface, vanishing exactly at the barrier. This
order-parameter profile induces the appearance of new bound states at the gap edges for
some trajectories \cite{Barash1997}, which are also visible in angle-averaged DOS, see Fig.~1a. 
As we show in Fig.~1a these gap singularities are not present in the non-self 
consistent calculation. This fact has important consequences in the I-V characteristics. 
As is shown in Ref.~[\onlinecite{Cuevas2001}], assuming a constant order parameter the
most prominent feature of the I-Vs is the appearance of SGS at $eV = \Delta/n$,
instead of in $eV = 2\Delta/n$ as in conventional superconductors. Its origin can be
understood as follows. Inside the gap the current is dominated by multiple Andreev
reflections (MAR). In this case there are two types of MAR processes: (a) those which 
connect the bound states with the gap edges and (b) the usual ones connecting the gap edges. 
The first ones give rise to the SGS at $eV = \Delta/n$, while the second could give rise to 
the series $eV = 2\Delta/n$. However, for the non-self consistent calculation the DOS at the 
gap edges is rather small (see Fig.~1a), and thus, at the opening of this second type of 
processes their probability is small. In the self-consistent case the main consequence of the 
presence of the gap singularities is the recovery of the odd terms in the SGS series $eV = 
2\Delta/n$. The odd terms appear now due to the enhancement of the probability of the MARs 
connecting the gap edges. However, the bound states at the gap edges do not appear for every 
trajectory, which weakens the SGS due to these processes. In this sense, in Fig.~\ref{45-45} one 
can clearly see a difference between the even and odd maxima of the conductance. This even-odd 
effect was reported in Ref.~[\onlinecite{Nesher1999}]. This result show clearly the relevance of 
the self-consistency, even in the ideal clean case.

\begin{figure}[t]
\begin{center}
\includegraphics[width=0.95\columnwidth,clip=]{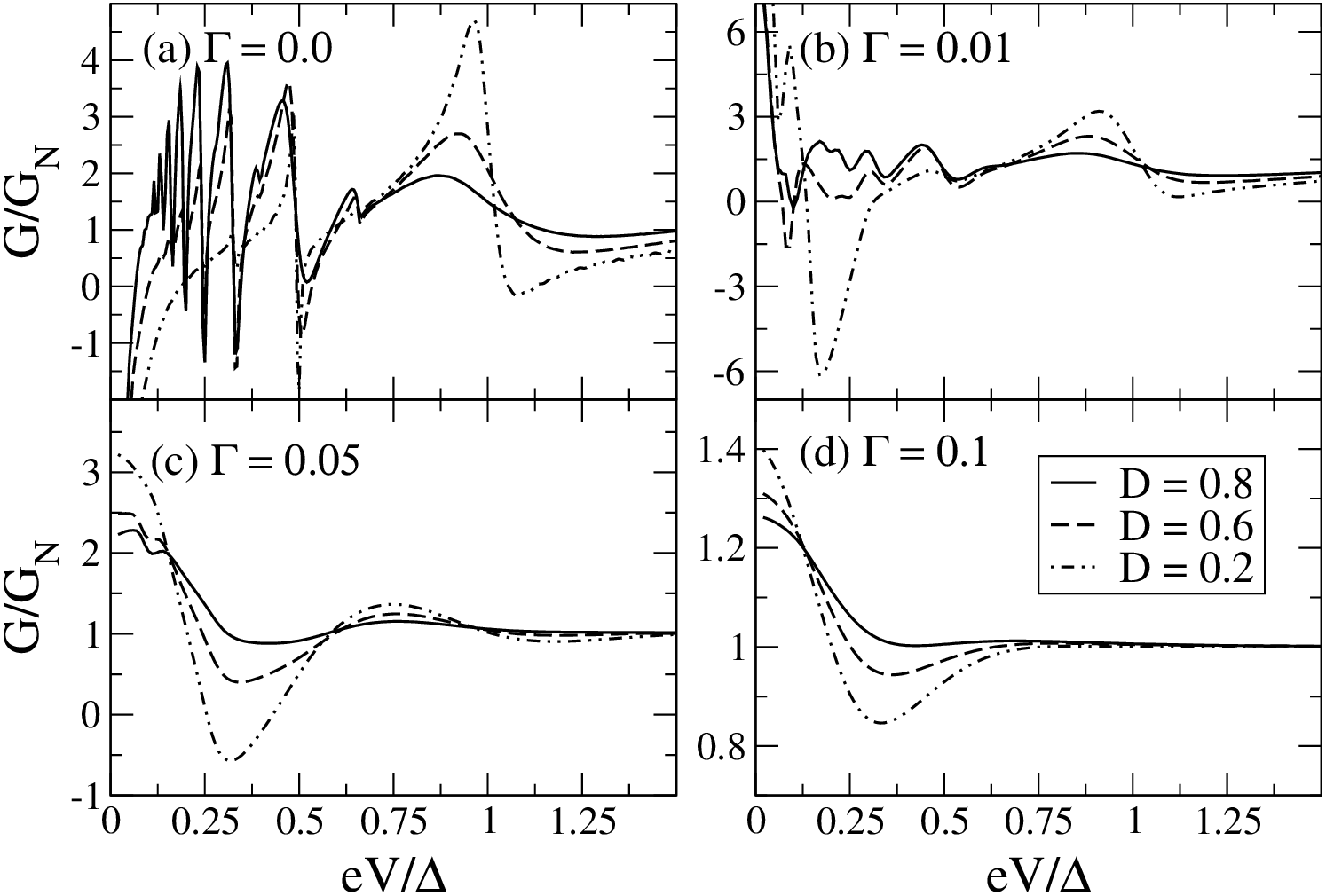}
\vspace*{-0.3truecm}
\caption{Differential conductance for a $d_{\pi/4}$-$d_{-\pi/4}$ contact for different values of the
bulk-impurity scattering rate $\Gamma$. Transmission rates as in  Fig.~2a.}
\label{impurities}
\end{center}
\end{figure}

In d-wave superconductors the order parameter is very sensitive to scattering from nonmagnetic
impurities and surface roughness. In particular, it is known that these elastic scattering
mechanisms provide an intrinsic broadening for the zero-energy bound states (ZEBS)
\cite{Poenicke1999,Fogelstrom-preprints}. For the case of Born scatterers this broadening is 
$\propto \sqrt{\Gamma \Delta}$, where $\Gamma=1/2\tau$ is the effective pair-breaking parameter 
locally at the surface. This is illustrated in Fig.~1b for the case of bulk impurities.
The natural question now is: what is the influence of the broadening of the ZEBS in
the SGS? In Fig.~\ref{impurities} we show the differential conductance of a $d_{\pi/4}$-$d_{-\pi/4}$ 
junction for different values of the bulk-impurity scattering rate. Notice that as the 
elastic scattering rate increases the SGS disappears, which can be understood as follows: the 
increase of density of states in the gap region enhances the probability of single-quasiparticle 
processes, producing the subsequent reduction of the probability of the Andreev processes, which 
in turn leads to the suppression of the SGS.

Probably the clearest signature of d-wave superconductivity in the SGS is its evolution
with magnetic field. It is known that a magnetic field perpendicular to the $ab$-plane,
$\vH=H\hat{\vz}$, leads to a Doppler shift in the continuum excitations given by 
$\vv_f\cdot\vp_s$, where the condensate momentum is $\vp_s = 
-(e/c)A(x)\hat{\vy}$, with $A$ the self-consistently determined vector potential 
\cite{Fogelstrom1997}. As shown in Fig.~1c, this means that the Andreev bound states are 
shifted to an energy which, in the limit of a large ratio $\lambda/\xi_0$, can be estimated 
to be $\epsilon_b(\pfhv) = (e/c)v_f H \lambda \sin \pfhv$, $\lambda$ being the $ab$-plane 
penetration depth. We shall use a natural field scale set by a screening current of order the 
bulk critical current, $H_0 = c\Delta/ev_f\lambda$, which is of the order of a Tesla 
\cite{Fogelstrom1997}. As can be seen in Fig.~1d, the screening currents flow 
parallel to the interface and in opposite directions in both electrodes, which means that the 
trajectory resolved DOS of the left and right superconductors are shifted by $2\epsilon_b(\pfhv)$ 
relative to each other. This shift modifies the threshold voltages of MARs starting and ending 
in different electrodes, leading to the splitting of the peaks with an odd order $n$ in the SGS. 
On the contrary, since the magnetic field produces a rigid shift of the spectrum, the threshold 
voltages of those MARs starting and ending in the same electrode are not modified. This means that 
the positions of the peaks with an even order $n$ in the SGS remain unchanged. This fact gives 
rise to a rich SGS in the trajectory resolved current, which consists in conductance peaks at the 
following voltage positions: (a) $[2\Delta(\pfhv) \pm 2\epsilon_b(\pfhv)]/n$, with $n$ odd, due 
to MARs connecting the gap edges of the left and right electrodes (the signs $\pm$ correspond to 
electron and hole processes), (b) $[\Delta(\pfhv) \pm 2\epsilon_b(\pfhv)]/n$, with $n$ odd, due 
to MARs connecting the bound states and the gap edges of different electrodes, (c) 
$2\Delta(\pfhv)/n$, with $n$ even, due to MARs connecting the gap edges of the same electrode, 
(d) $\Delta(\pfhv)/n$, with $n$ even, due to MARs connecting the bound states and the gap edges 
of the same electrode. Finally, there is also structure at $eV=2\epsilon_b(\pfhv)/n$, with $n$ odd, 
due to MARs connecting the bound states.

\begin{figure}[t]
\begin{center}
\includegraphics[width=.9\columnwidth,clip=]{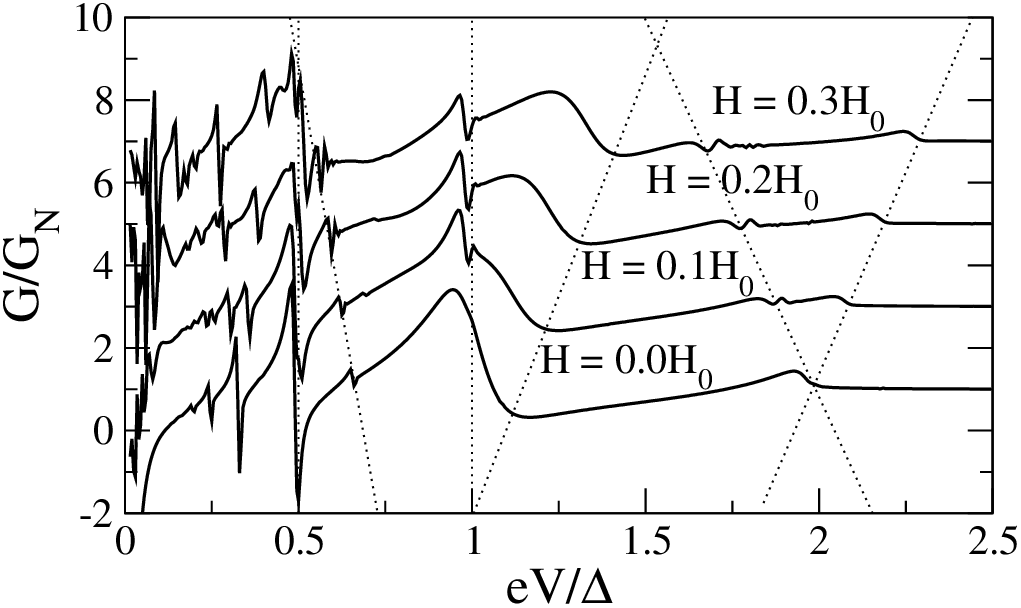}
\vspace*{-0.3truecm}
\caption{Differential conductance of a clean  $d_{\pi/4}$-$d_{-\pi/4}$ junction with $D=0.4$
and different values of the magnetic field. The curves has been vertically displaced for
clarity and dotted lines have been added to guide the eye.}
\label{B-field}
\end{center}
\end{figure}

After doing the angle average most of these features are still clearly visible. This is illustrated 
in Fig.~\ref{B-field} where we show the differential conductance of a $d_{\pi/4}$-$d_{-\pi/4}$ 
junction with transmission $D=0.4$ for different values of the magnetic field. Starting at large 
voltages, the weak structure seen at $2\Delta$ splits with applied field. Around $eV=\Delta$ there 
are both type (b) and type (c) processes leading to a maximum at $eV=\Delta$, unaffected by the 
applied field, as well as a field-shift of the dip just above $\Delta$. The field dependence of the 
differential conductance is most clearly resolved at larger biases, $eV\ge \Delta/2$, as the marks of 
the various processes begin to overlap at small bias. What is important to note is that it is still the 
bulk maximum gap that gives the characteristic energy scale for the SGS. The effect of the Doppler 
shift on the SGS or the differential conductance is only prominent in junctions with a sizable 
misorientation. For junctions close to the $d_{0}$-$d_{0}$ case, the main contribution to the SGS 
comes from trajectories close to perpendicular incidence, i.e. with $\sin \pfhv \sim 0$ and thus 
having a vanishing Doppler shift.

In summary, we have presented a self-consistent analysis of the I-V characteristics of d-wave/d-wave 
contacts at high transparencies. We have shown that it is possible to observe SGS for all crystal 
misorientations. We predict that the presence of bound states inside the gap gives rise to two 
new qualitative effects in the SGS: (a) an even-odd effect, and (b) a very peculiar splitting in an 
external magnetic field. These features are unique hallmarks of the d-wave scenario, and we hope 
that our analysis will trigger off a more detailed experimental study of the SGS in cuprate 
Josephson junctions.

This work has been supported by the EU TMR Network on Dynamics of Nanostructures, the CFN supported 
by the DFG, and by the Swedish research council.


\end{document}